\documentclass[superscriptaddress,aps,prb,amsmath,amssymb,twocolumn,letterpaper,
         footinbib,bibnotes,showpacs,reprint,balancelastpage,raggedbottom,longbibliography,%
         citeautoscript,floatfix
]{revtex4-2}

\usepackage{placeins}
\usepackage{graphicx}
\usepackage{subcaption}
\usepackage{float}
\usepackage{dcolumn}
\usepackage{bm}
\usepackage[dvipsnames,table]{xcolor}
\usepackage[%
  bookmarks = false, %
  breaklinks,             %
  pdfpagemode   = UseNone,%
  pdfstartview  = FitH,     %
  pdfstartpage  = 1,      %
  colorlinks    = true   %
]{hyperref}
\hypersetup{
    linkcolor=RubineRed,          
    citecolor=RoyalBlue,        
    filecolor=Mulberry,      
    urlcolor=RoyalBlue           
}
\usepackage{amssymb}
\usepackage{amsmath}
\usepackage{float}
\usepackage{enumerate}
\usepackage{microtype}
\usepackage{xspace}
\usepackage{siunitx}
\usepackage{upgreek}
\usepackage{nicefrac}
\usepackage{booktabs}
\usepackage{soul}  
\usepackage[none]{hyphenat} 
\usepackage{hyperref}
\usepackage{xr-hyper}



\begin{document}

\preprint{AIP/123-QED}

\title{Defect Chemistry of Hafnia-Based Ferroelectrics under Oxide-Electrode Constraints}

\author{Sehyeok Jeon}
\affiliation{Department of Materials Science and Engineering, Yonsei University}

\author{Aloysius Soon}
\email{aloysius.soon@yonsei.ac.kr}
\affiliation{Department of Materials Science and Engineering, Yonsei University}

\author{Yongjin Shin}
\email{yongjin.shin@dankook.ac.kr}
\affiliation{Department of Semiconductor Convergence Engineering, Dankook University}

\date{\today}

\begin{abstract}

HfO$_2$ heterostructures with La$_{0.67}$Sr$_{0.33}$MnO$_3$ (LSMO) epitaxial electrodes have attracted interest as ferroelectric systems with electrode-dependent defect chemistry.
However, the formation and stability of defects under LSMO-constrained chemical conditions remain poorly understood.
In this work, we use density functional theory calculations to systematically investigate oxygen vacancies and La-, Sr-, and Mn-related interstitial and substitutional defects in HfO$_2$ under chemical environments constrained by the LSMO electrode. We find that LSMO stability substantially narrows the accessible O chemical potential range relative to bulk HfO$_2$, altering the predicted defect formation energies. While La and Sr substitutional defects are governed primarily by preferred ionic valence states, oxygen vacancies and Mn substitutional defects exhibit charge states sensitive to the local atomic environment. In addition, defect-complex calculations in orthorhombic HfO$_2$ reveal negative binding energies, indicating favorable defect association.
The La- and Mn-related defect complexes exhibit thermodynamic competition, with their relative stability depending on the Fermi level and oxygen chemical potential.
Lastly, extending the analysis to Hf$_{0.5}$Zr$_{0.5}$O$_2$ and ZrO$_2$ reveals common oxygen-vacancy trends, while the stability of LSMO-derived defects depends more strongly on the host material and crystal phase. These results provide a thermodynamic basis for understanding how electrode-constrained chemical environments and host structures influence defect formation, charge compensation, and defect-complex stability in HfO$_2$-based materials.
%
\end{abstract}

\maketitle

\section{Introduction}


HfO$_2$-based materials have attracted considerable attention owing to their compatibility with complementary metal-oxide-semiconductor (CMOS) technology and the emergence of ferroelectricity in the polar orthorhombic (o) phase~\cite{092_Cheol_Seong_Hwang_2022,093_M.Acuautla_2023,063_S.Liu_2024,058_Jingsheng.Chen_2026,008ShiyouChen2021}.
However, the polar o-phase is metastable under bulk equilibrium, making its stabilization challenging~\cite{008ShiyouChen2021,060_NS.Fedorova_2024}.
The relative stability of HfO$_2$ polymorphs is influenced by various factors, including composition, strain, interfaces, and defects~\cite{007J_Van_Houdt2021,008ShiyouChen2021,060_NS.Fedorova_2024,063_S.Liu_2024}. In particular, oxygen vacancies ($V_{\text{O}}$) and incorporated impurities can modify local bonding, charge distributions, and lattice distortions, influencing the relative stability of different crystalline phases~\cite{008ShiyouChen2021,058_Jingsheng.Chen_2026,060_NS.Fedorova_2024,025_MinLiao_2026,050_Min.Liao_2023,018_Satishchandra_Ogale_2021,016_Florencio_Sanchez_2021,Jeon2026}.

Electrodes in contact with HfO$_2$-based films can influence both phase stability and defect behavior by modifying the structural and chemical environment  \cite{013JunMingLiu_2022,015_Florencio_Sanchez_2021, 016_Florencio_Sanchez_2021}.
In particular, La$_{0.67}$Sr$_{0.33}$MnO$_3$ (LSMO) is widely used as a bottom electrode and epitaxial template for HfO$_2$ and Hf$_{0.5}$Zr$_{0.5}$O$_2$ (HZO) thin films \cite{058_Jingsheng.Chen_2026,016_Florencio_Sanchez_2021}. Substrate-induced tensile strain in LSMO favors o-phase formation and enhances HZO polarization \cite{017_Florencio_Sanchez_2019}, while the chemical composition and atomic structure of the LSMO surface have been proposed to further favor o-HZO by lowering the interface energy relative to competing polymorphs \cite{016_Florencio_Sanchez_2021}. Beyond these structural effects, changes in the Mn$^{3+}$/Mn$^{4+}$ ratio in LSMO under an applied electric field can accompany oxygen-vacancy migration and redistribution near the HfO$_2$/LSMO interface~\cite{018_Satishchandra_Ogale_2021}.

The constituent cations of LSMO, La, Sr, and Mn, have also been shown experimentally to influence the formation and stabilization when introduced as dopants in HfO$_2$~\cite{071_Thomas.Mikolajick_2018,072_D.Zhang_2018,091_Jing.Ma_2024}.
Experimental and theoretical studies further connect phase stability and electrical behavior in these doped systems to oxygen-vacancy content, vacancy charge states, and charge compensation involving cation defects~\cite{050_Min.Liao_2023,049_Alfred.Kersch_2017,088_Zuhuang.Chen_2024}.
These findings suggest that understanding the effects of cation incorporation requires considering oxygen vacancies and cation defects together.
For HfO$_2$ in thermodynamic equilibrium with LSMO, the stability of both materials constrains the chemical potentials governing the formation of these defects.
However, the formation energies and preferred charge states of these defects have not been systematically compared under the chemical-potential constraints imposed by LSMO stability.
A systematic comparison of their formation energetics and charge states across the Fermi-level range is therefore needed to clarify the possible charge-compensation mechanisms in HfO$_2$ under LSMO-constrained chemical conditions.

In this work, we use first-principles calculations to investigate oxygen vacancies and La-, Sr-, and Mn-related interstitial and substitutional defects in o-HfO$_2$, with LSMO considered as a possible source of the cation defects. We first determine the accessible O chemical-potential range by accounting for the thermodynamic stability of LSMO and find that it is narrower than that permitted by the host oxide alone. Defect formation-energy diagrams identify the thermodynamically stable charge states and show that oxygen-vacancy site preferences depend on the charge state and local cation environment. La and Sr substitutional defects favor negative charge states that compensate the electron deficiencies introduced by replacing Hf$^{4+}$ with La$^{3+}$ and Sr$^{2+}$.
In contrast, Mn substitutional defects show charge-dependent electronic and magnetic responses involving both Mn and neighboring O atoms.
In addition, defect-complex calculations reveal negative binding energies, indicating energetically favorable interactions between substitutional defects and oxygen vacancies. The relative stability of La- and Mn-containing complexes depends on the Fermi-level position and chemical-potential conditions, revealing thermodynamic competition between these complexes. Lastly, we extend the analysis to other HfO$_2$ phases and to HZO and ZrO$_2$ to examine how structural and chemical environments influence defect behavior. Our insights provide a basis for understanding how electrode chemistry and host structure jointly influence defect formation in HfO$_2$-based ferroelectrics.

\section{computational methods}


We performed density functional theory (DFT) calculations \cite{074_Kohn_1965} using the Vienna \textit{Ab Initio} Simulation Package (VASP) \cite{075_G.Kresse_1996}.
The projector augmented wave (PAW) method was employed to describe the valence electrons using the following PAW potentials: Hf\_pv ($5p^{6}5d^{4}6s^{1}$), Zr\_sv ($4s^{2}4p^{6}4d^{3}5s^{1}$), O ($2s^{2}2p^{4}$), La ($4f^{0.0001}5s^{2}5p^{6}5d^{0.999}6s^{2}$), Sr\_sv ($4s^{2}4p^{6}4d^{0.001}5s^{1.999}$), and Mn\_pv ($3p^{6}3d^{6}4s^{1}$)~\cite{076_G.Kresse_1999,077_PE.Blochi_1994}. The crystal structures and competing phases used to determine the chemical-potential bounds were first optimized using the Perdew--Burke--Ernzerhof functional revised for solids (PBEsol) \cite{081_Kieron.Burke_2008}.
The competing phases were identified using the \texttt{doped} code based on data from the Materials Project \cite{038_David_O.scanlon}, from which only the crystal structures were adopted.
For Mn-containing systems, the PBEsol+$U$ method \cite{080_A.P.Sutton_1998} was employed with $U_{\mathrm{eff}} = 3$~eV to describe the localized Mn $3d$ orbitals, consistent with previous works on manganite perovskites \cite{082_Shin/Rondinelli2020,083_Shin/Rondinelli2021}.
Subsequently, single-point calculations were performed on the PBEsol-relaxed structures using the Heyd--Scuseria--Ernzerhof hybrid functional (HSE06) \cite{078_Matthias.Ernzerhof_2003,079_Gustavo.E.Scuseria_2006}, and the resulting total energies were used to calculate defect formation energies and chemical-potential bounds.
Electronic structures were evaluated using the HSE06 to address band-gap underestimation by semilocal functionals.
A plane-wave energy cutoff of 500~eV was used, with a $2 \times 2 \times 2$ $k$-point mesh for PBEsol calculations and $\Gamma$-point sampling for HSE06 calculations.

Cell parameters and atomic positions were optimized until all atomic forces were below $0.01$~eV/\AA.
We used $2 \times 2 \times 2$ supercells for the monoclinic (m) and o-phases and a $2 \times 2 \times 1$ supercell for the antipolar orthorhombic (ao) phase, each containing 96 atoms.
The tetragonal (t) phase was modeled using a 108-atom $3 \times 3 \times 2$ supercell.
Defect-free HfO$_2$, ZrO$_2$, and HZO structures were fully relaxed, while only atomic positions were relaxed in defect-containing supercells with lattice parameters fixed at the optimized bulk values.
For interstitial defects, initial candidate sites were generated using voronoi tessellation, followed by screening based on crystal symmetry to eliminate symmetrically equivalent sites~\cite{038_David_O.scanlon}.
All selected candidate structures were then structurally relaxed, and the lowest-energy configuration was used for the defect formation-energy calculations.
The HSE06 functional yielded band gaps of 5.65 and 6.2 eV for the m- and o-phases, respectively, in good agreement with the previously reported values of 5.6 \cite{008ShiyouChen2021} and 6.11 eV \cite{066_Xiangshui.Miad_2023,067_Jibao.Wu_2021}.

The formation energy of a defect $D$ in charge state $q$ was calculated as~\cite{022_ChrisG.VandeWalle_2014,023_G.Galli_2025,Shin/VdW2025} 
\begin{equation}
\begin{aligned}
E^q[D^q] &=  E_{\mathrm{tot}}[D^q] - E_{\mathrm{tot}}[\mathrm{bulk}] \\
&- \sum_i n_i\mu_i + q(\varepsilon_{\mathrm{F}} + E_{\mathrm{VBM}}) + E_\mathrm{corr},
\end{aligned}
\label{eq:DFE}
\end{equation}
where $E_{\mathrm{tot}}[D^q]$ and $E_{\mathrm{tot}}[\mathrm{bulk}]$ are the total energies of the defective and defect-free supercells, respectively.
$n_i$ counts the atoms of species $i$ added ($n_i > 0$) or removed ($n_i < 0$), and $\mu_i$ is the corresponding chemical potential.
The Fermi level $\varepsilon_{\mathrm{F}}$ is referenced to the valence-band maximum ($E_{\mathrm{VBM}}$), and $E_{\mathrm{corr}}$ is the finite-size correction evaluated using the Freysoldt--Neugebauer--Van de Walle (FNV) scheme \cite{021_ChrisG.VandeWalle_2009}. \autoref{tab:dielectric} presents the dielectric constants of HfO$_2$, ZrO$_2$, and HZO for the different crystal phases \cite{020_A.Kersch_2015}.

We also evaluated the binding energies of defect complexes, which are independent of the chemical potentials, to assess the interactions between substitutional defects
and oxygen vacancies. The binding energy was defined as~\cite{094_Richard_G.Hennig_2020} 

\begin{equation}
\begin{aligned}
\Delta E =
E_f[(nD_{\mathrm{Hf}}-V_{\mathrm O})^{q''}]
-nE_f[D_{\mathrm{Hf}}^{q}]
-E_f[V_{\mathrm O}^{q'}],
\end{aligned}
\label{eq:binding}
\end{equation}

where $D_{\mathrm{Hf}}$ represents a substitutional defect, $n$ is the number of substitutional defects ($n=1$ or $2$), and $q$, $q'$, and $q''$ denote the charge states of the substitutional defect, oxygen vacancy, and defect complex,
respectively.
The charge states were chosen to satisfy $q''=nq+q'$, ensuring that the binding energies are independent of the Fermi level. A negative $\Delta E$ indicates that the formation of the defect complex is energetically favorable relative to the isolated defects.

\begin{table}[htbp]
\centering
\caption{Space groups and dielectric constants of HfO$_2$, ZrO$_2$, and HZO for different crystal phases.}
\label{tab:dielectric}
\begin{ruledtabular}
\begin{tabular}{l|lccc}
Phase & Space group & HfO$_2$ & ZrO$_2$ & HZO \\
\hline
o-phase      & $Pca2_1$ (No. 29) & 27.0 & 31.3 & 29.1  \\
ao-phase & $Pbca$ (No. 61) & 27.0 & 31.3 & 29.1 \\
m-phase              & $P2_1/c$ (No. 14) & 22.4 & 25.8 & 24.1  \\
t-phase              & $P4_2/nmc$ (No. 137) & 45.9 & 49.4 & 48.7  \\
\end{tabular}
\end{ruledtabular}
\end{table}

\begin{figure*}[t]
  \centering
  \includegraphics[width=0.85\linewidth,trim=0 0cm 0 0cm,clip]{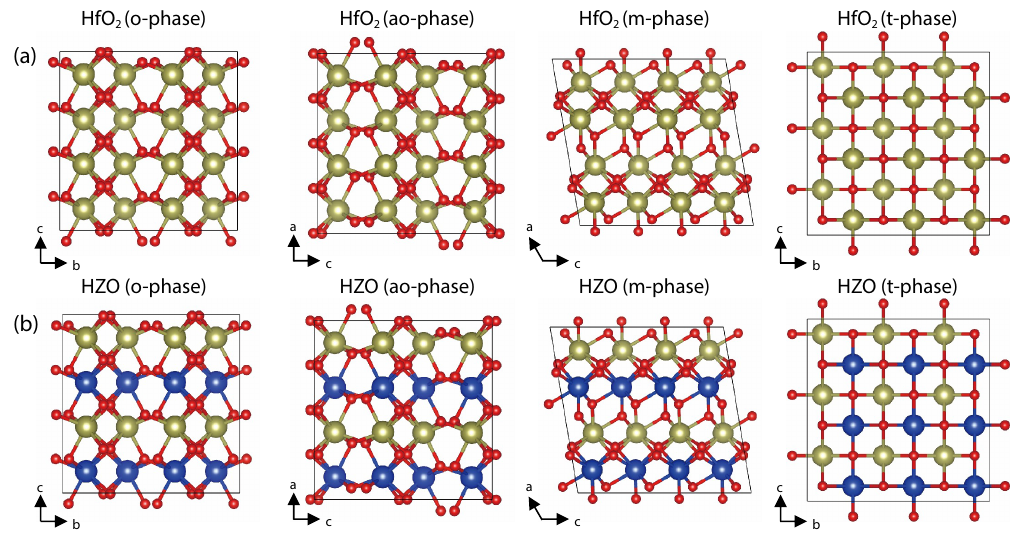}
  \caption{The crystal structures of the o-, ao-, m-, and t-phases of (a) HfO$_2$ and (b) Hf$_{0.5}$Zr$_{0.5}$O$_2$ (HZO) with layered Hf/Zr ordering. Hf, Zr, and O atoms are represented by olive, blue, and red, respectively.
  }
  \label{fig:All_structures}
\end{figure*}

\begin{figure}[t]
  \centering
  \includegraphics[width=0.95\linewidth,trim=0 0cm 0 0cm,clip]{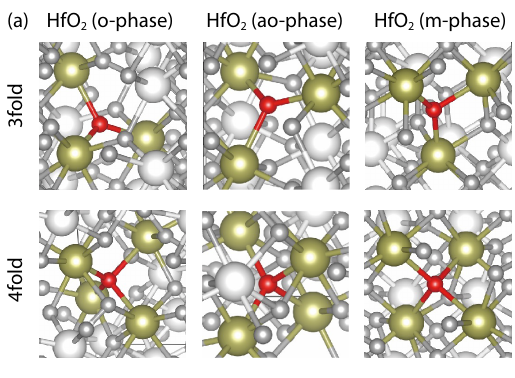}
  \caption{Local coordination environments of the threefold-coordinated (3fold) and fourfold-coordinated (4fold) oxygen sites in the o-, ao-, and m-phases of HfO$_2$. The selected O atoms and their nearest-neighbor Hf atoms are shown in red and olive, respectively; all other atoms are shown in gray for clarity.
  }
  \label{fig:Local_structures}
\end{figure}
\section{Results and discussion}

\subsection{Atomic structure of HfO$_2$ phases}

HfO$_2$ exhibits several polymorphs, including the ferroelectric o-, m-, ao-, and t-phases~\cite{008ShiyouChen2021,063_S.Liu_2024}. We consider these phases because the m-phase is the bulk ground state under ambient conditions~\cite{007J_Van_Houdt2021,008ShiyouChen2021}, while the ao- and t-phases have been identified as possible intermediates along ferroelectric switching pathways~\cite{031_YoshioNishi_2017,032_JiezhiChen_2024,063_S.Liu_2024}.
We focus primarily on o-phase defect properties and examine the other three phases later.

\autoref{fig:All_structures} shows the four surveyed phases of HfO$_2$ and HZO, with the corresponding space groups summarized in \autoref{tab:dielectric}.
Among the phases examined, t-HfO$_2$ has the highest structural symmetry, with all Hf sites crystallographically equivalent to one another and all O sites likewise equivalent.
In this phase, each Hf atom is coordinated by eight O atoms, and each O atom is coordinated by four Hf atoms.
The o-, ao-, and m-phases exhibit distortions of the Hf--O framework relative to the t-phase, resulting in lower crystal symmetry and changes in local coordination~\cite{007J_Van_Houdt2021,008ShiyouChen2021,034_DavidVanderbilt_2002}.
Specifically, the Hf coordination number decreases from eight to seven, while the O atoms occupy two crystallographically inequivalent sites: threefold-coordinated (3fold) and fourfold-coordinated (4fold), as illustrated in \autoref{fig:Local_structures}.

For HZO, we adopted layered Hf/Zr ordering, as previous studies reported relatively small differences in total energy and other relevant properties between layered and alternating cation arrangements~\cite{020_A.Kersch_2015,035_KarinM.Rabe_2020}. Our calculations also showed a total energy difference between the two arrangements was only 0.5~meV/f.u. for both the ao- and m-phases. This weak dependence on cation ordering is consistent with the similar chemical and physical properties of Hf and Zr. Both elements favor the $+4$ oxidation state, with atomic radii of 1.44 and 1.45~\AA{}, ionic radii of 0.85 and 0.86~\AA{}, and electronegativities of 1.23 and 1.22 for Hf and Zr, respectively~\cite{012_WK_Choi_2007}. Hf/Zr mixing additionally introduces different local cation environments for oxygen sites with the same coordination number. For example, threefold oxygen sites with nearest-neighbor cation environments (Hf, Hf, Zr) and (Zr, Zr, Hf) are denoted as 3fold (Hf) and 3fold (Zr), respectively.

\subsection{Evaluation of chemical potential}
\begin{figure*}[t]
  \centering
  \includegraphics[width=0.95\linewidth,trim=0 0cm 0 0cm,clip]{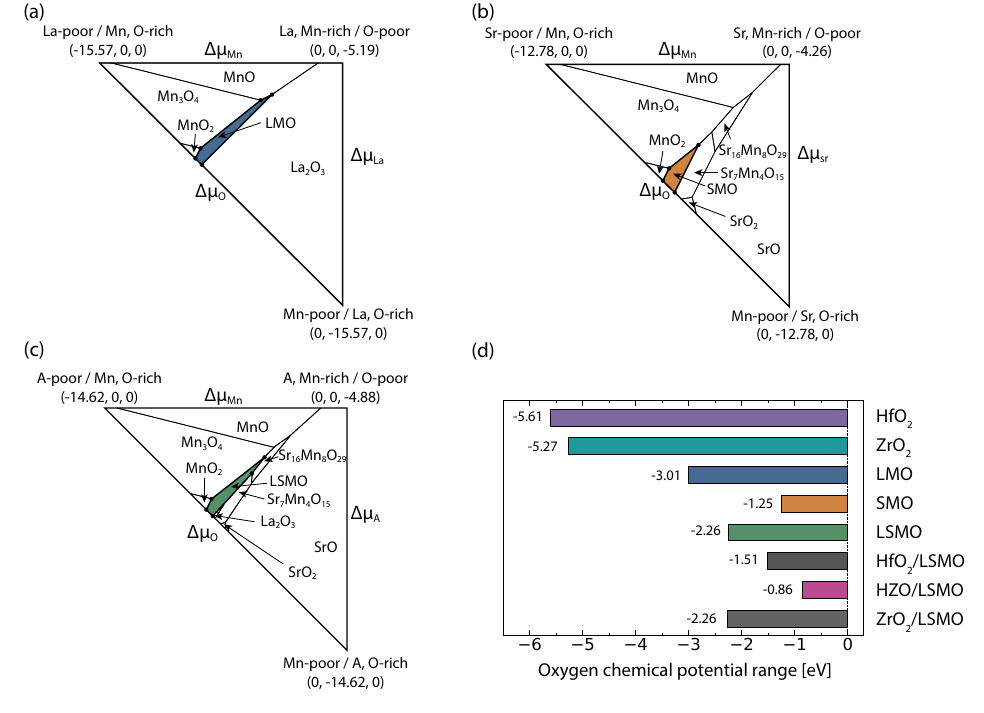}
  \caption{Accessible chemical potential ranges for (a) LMO, (b) SMO, and (c) LSMO. The values in parentheses indicate the accessible chemical potential ranges of the corresponding phases composed of the constituent elements. Each side of the triangle represents the chemical potential boundary under the element-rich condition for the corresponding element. For LSMO, $A$ is defined as the weighted chemical potential of the A-site cations, $A=\frac{2}{3}\Delta\mu_{\mathrm{La}}+\frac{1}{3}\Delta\mu_{\mathrm{Sr}}$, for visualization in the two-dimensional chemical potential diagram. (d) Accessible range of the O chemical potential, $\Delta\mu_{\mathrm{O}}$, determined by the thermodynamic stability of each host material.
  }
  \label{fig:Chemical_potential}
\end{figure*}

The chemical potential accounts for the energy change in an external reservoir associated with the exchange of atomic species during defect formation. Experimental conditions constrain the chemical potentials and thus affect the calculated defect formation energies through \autoref{eq:DFE}~\cite{022_ChrisG.VandeWalle_2014}.

For HfO$_2$ in thermodynamic equilibrium, the chemical potentials satisfy
\begin{equation}
\mu_{\mathrm{Hf}} + 2\mu_{\mathrm{O}} = \mu_{\mathrm{HfO_2}}.
\label{eq:hfo2_equilibrium}
\end{equation}
The chemical potential of each element can be expressed relative to the corresponding elemental reference state as
\begin{equation}
\mu_i = \mu_i^{0} + \Delta\mu_i,
\label{eq:chemical}
\end{equation}
where $\mu_i^{0}$ is the chemical potential of element $i$ in the elemental reference state, and $\Delta\mu_i$ is the deviation from this reference.
Substituting \autoref{eq:chemical} into \autoref{eq:hfo2_equilibrium} gives the host equilibrium condition in terms of these deviations:
\begin{equation}
\Delta\mu_{\mathrm{Hf}} + 2\Delta\mu_{\mathrm{O}}
=
\Delta H_f(\mathrm{HfO_2}),
\end{equation}
where $\Delta H_f(\mathrm{HfO_2})$ is the formation energy of HfO$_2$~\cite{022_ChrisG.VandeWalle_2014}.

To prevent the formation of elemental Hf and O$_2$, the chemical potentials must satisfy
\begin{equation}
\Delta\mu_{\mathrm{Hf}} \leq 0,
\qquad
\Delta\mu_{\mathrm{O}} \leq 0.
\end{equation}

The thermodynamically allowed chemical potential range is determined by the host equilibrium condition and elemental precipitation constraints.
Within this range, the O-rich limit corresponds to the maximum accessible $\Delta\mu_{\mathrm{O}}$, typically zero, whereas the O-poor limit is the lowest value consistent with HfO$_2$ stability.
The same procedure was applied to ZrO$_2$ and HZO to determine the corresponding chemical-potential limits for evaluating defect formation energies.

We extended this approach to LMO, SMO, and LSMO, including additional constraints imposed by competing phases.
For LSMO, the host equilibrium condition and elemental precipitation constraints are
\begin{equation}
\begin{array}{c}
\displaystyle
\frac{2}{3}\Delta\mu_{\mathrm{La}}
+\frac{1}{3}\Delta\mu_{\mathrm{Sr}}
+\Delta\mu_{\mathrm{Mn}}
+3\Delta\mu_{\mathrm{O}}
= \Delta H_f(\mathrm{LSMO})
\\[6pt]

\displaystyle
\Delta\mu_{\mathrm{La}},\;
\Delta\mu_{\mathrm{Sr}},\;
\Delta\mu_{\mathrm{Mn}},\;
\Delta\mu_{\mathrm{O}}
\leq 0.
\end{array}
\label{eq:LSMO_stability}
\end{equation}
To prevent the formation of competing phases, the stoichiometrically weighted sum of the constituent chemical potential deviations must not exceed the formation enthalpy of each competing phase. The competing phases and corresponding constraints are listed in Section~S3.

\autoref{fig:Chemical_potential}(a-c) show the stability regions of LMO, SMO, and LSMO, together with the regions where competing phases become stable and constrain host stability.
To visualize the chemical-potential stability region of quaternary LSMO in \autoref{fig:Chemical_potential}(c), we define the weighted $A$-site chemical potential as $\Delta\mu_{\mathrm{A}}=\frac{2}{3}\Delta\mu_{\mathrm{La}}+\frac{1}{3}\Delta\mu_{\mathrm{Sr}}$.
Within the allowed chemical-potential ranges, we evaluated defect formation energies under O-rich and O-poor conditions because the oxygen chemical potential is a key variable during synthesis~\cite{090_Morgan.Trassin_2023}.
We additionally assumed Mn-rich conditions, motivated by previous reports of Mn segregation~\cite{036_N.Pinna_2009,037_RachidAhlLaamara_2025}.
Finally, we adopted La-rich/Sr-poor conditions, with La and Sr chemical potentials differing by 0.38 and 0.76~eV, respectively, from those under La-poor/Sr-rich conditions.

As shown in \autoref{fig:Chemical_potential}(d), the accessible oxygen chemical-potential range depends strongly on which materials are included in the thermodynamic stability analysis.
For HfO$_2$ and ZrO$_2$ considered separately, the O-poor limits are $-5.61$ and $-5.27$~eV, respectively~\cite{008ShiyouChen2021}.
When only the LSMO electrode is considered, the O-poor limit is $-2.26$~eV.
Applying the combined stability constraints for HfO$_2$/LSMO raises the O-poor limit to $-1.51$~eV, substantially narrowing the accessible range compared with HfO$_2$ alone.
Similar restrictions are also observed for HZO/LSMO and ZrO$_2$/LSMO.
These results highlight the importance of including electrode stability constraints when selecting chemical potentials for defect formation calculations.

\subsection{Oxygen-vacancy properties in orthorhombic HfO$_2$}

\begin{figure*}[t]
  \centering
  \includegraphics[width=0.65\textwidth,trim=0 0cm 0 0cm,clip]{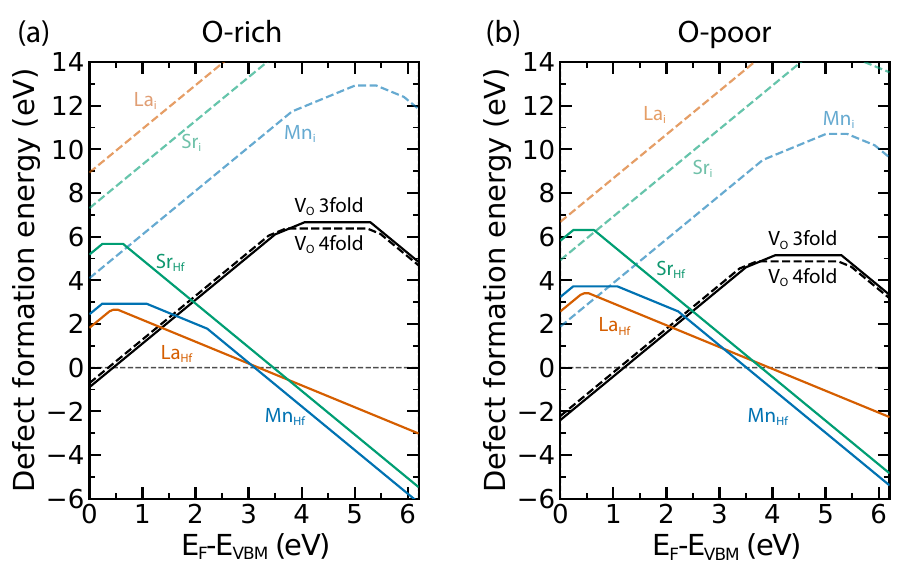}
  \caption{Defect formation energy diagrams of oxygen vacancies and La, Sr, and Mn interstitial and substitutional defects in o-HfO$_2$ under (a) O-rich and (b) O-poor conditions.}
  \label{fig:DFE_HFO}
\end{figure*}

In \autoref{fig:DFE_HFO}, we show the calculated formation energies of oxygen vacancies and La-, Sr-, and Mn-related interstitial and substitutional defects in o-HfO$_2$, using the chemical potentials established above. We first discuss oxygen vacancies, then LSMO-derived defects, and finally extend the analysis to other phases of HfO$_2$ and to HZO and ZrO$_2$.

Under O-rich conditions ($\Delta\mu_{\mathrm{O}}=0$), vacancies at 3fold and 4fold oxygen sites exhibit similar formation-energy trends, although the relative stability depends on the charge state.
Specifically, the 3fold vacancy is more stable in positively charged states, whereas the 4fold vacancy is more stable in the neutral and negatively charged states, consistent with previous studies~\cite{007J_Van_Houdt2021,008ShiyouChen2021}.
The formation-energy difference between the two sites is 0.28~eV in the neutral state, compared with 0.18~eV in both the $q=-2$ and $q=+2$ states.
These differences may reflect changes in the local Hf bonding environment upon oxygen removal.
Formation of $V_{\text{O}}^{0}$ removes the original bonding states at both sites, while an additional occupied antibonding state at the 3fold site raises the formation energy~\cite{089_Giuliana.Di.Martino_2026}.
In the $q=+2$ state, this antibonding state lies above the Fermi level and remains unoccupied, reducing the formation-energy difference between the two sites compared with the neutral state~\cite{089_Giuliana.Di.Martino_2026}.
The charge transition levels (CTLs) differ by at most 0.35~eV between the two vacancy sites, as summarized in \autoref{tab:CTL_VO}. The largest difference occurs for the $(+1/0)$ transition, reflecting the lower stability of the $q=+1$ state relative to the neutral state at the 4fold site.
For both configurations, the formation energy of $V_{\mathrm{O}}^{2+}$ becomes negative when the Fermi level is close to the VBM, indicating energetically favorable vacancy formation even under O-rich conditions.

\begin{figure*}[t]
  \centering
  \includegraphics[width=0.95\linewidth,trim=0 0cm 0 0cm,clip]{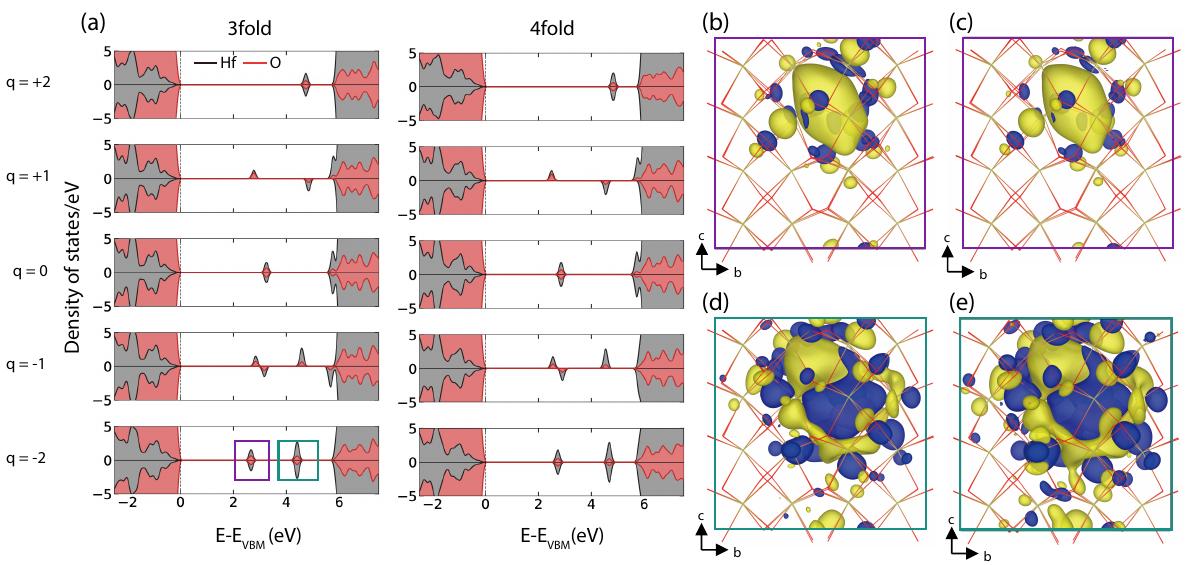}
  \caption{(a) Spin-resolved projected density of states (PDOS) of 3fold and 4fold sites in o-HfO$_2$ for charge states from $q=+2$ to $q=-2$. 
(b)--(e) Wavefunctions corresponding to the defect state peaks highlighted by the colored boxes in (a) for the 3fold site in the $q=-2$ charge state. (b) and (c) show the spin-up and spin-down wavefunctions, respectively, corresponding to the peak highlighted by the purple box, while (d) and (e) show the spin-up and spin-down wavefunctions, respectively, corresponding to the peak highlighted by the teal box. In each wavefunction, yellow and blue represent the positive and negative phases, respectively.
}
  \label{fig:DOS_VO}
\end{figure*}

As shown in \autoref{fig:DOS_VO}(a), neutral oxygen vacancies at the 3fold and 4fold sites produce deep, doubly occupied in-gap states at similar energies, approximately 3~eV above the VBM. These defect states derive from the dangling orbitals of the undercoordinated Hf atoms neighboring $V_{\text{O}}$~\cite{050_Min.Liao_2023}.
The occupation of these defect states also influences the structural relaxation of the neighboring Hf atoms. In $V_{\text{O}}^{0}$, the two electrons occupying the defect state screen the electrostatic interactions among the neighboring Hf ions.
Removing one electron to form $V_{\text{O}}^{+1}$ reduces this screening, causing the neighboring Hf atoms to relax outward. Relative to the positions in the defect-free structure, the neighboring Hf atoms move away from the original O site by an average of 0.109 and 0.090~\AA{} for the 3fold and 4fold vacancies, respectively~\cite{051_S.J.Clark_2006,050_Min.Liao_2023,085_A.L.Shluger_2007}.
Removing the second electron to form $V_{\text{O}}^{2+}$ further reduces the screening, increasing the average outward displacements to 0.223 and 0.211~\AA{} for the 3fold and 4fold sites, respectively.
This outward relaxation accompanies an upward shift of the vacancy-related defect state toward the CBM, as shown in \autoref{fig:DOS_VO}(a)~\cite{051_S.J.Clark_2006}.

In contrast, adding electrons to $V_{\text{O}}^{0}$ slightly lowers the energy of the pre-existing doubly occupied deep defect state and introduces an additional state near the CBM. The additional state appears to originate from a resonant state within the conduction band in the positively charged configurations, which shifts below the CBM upon electron addition~\cite{085_A.L.Shluger_2007}.
Changes in the electronic structure are accompanied by inward relaxation of the neighboring Hf atoms toward the vacancy in the negatively charged states.
Relative to the defect-free structure, the average inward displacements of the neighboring Hf atoms at the 3fold site are 0.057 and 0.132~\AA{} for $V_{\text{O}}^{-1}$ and $V_{\text{O}}^{-2}$, respectively. The corresponding displacements at the 4fold site are 0.065 and 0.117~\AA{}. For $V_{\text{O}}^{-2}$ at the 3fold site, the deeper defect states are strongly localized around the vacancy site [\autoref{fig:DOS_VO}(b,c)], whereas the states closer to the CBM extend farther over the surrounding atoms [\autoref{fig:DOS_VO}(d,e)].
The broader spatial distribution is consistent with the proposed origin of these states in conduction-band resonances~\cite{085_A.L.Shluger_2007}.

The relationship between Hf displacement and the additional state near the CBM is particularly evident for the 3fold $V_{\text{O}}^{-1}$. Projected DOS analysis shows that the additional state immediately below the CBM mainly originates from two neighboring Hf atoms.
When these Hf atoms were progressively moved toward the corresponding positions in the defect-free structure, the associated DOS peak gradually disappeared, suggesting that local Hf relaxation helps stabilize the additional state below the CBM.
The corresponding wavefunction extends over the surrounding atoms, as shown in Section S5. However, since the $q=-1$ charge state has almost no thermodynamically stable Fermi level range, the thermodynamic relevance of this defect state is limited. 

Under O-poor conditions, the oxygen-vacancy formation-energy curves undergo a rigid downward shift of approximately 1.5~eV relative to the O-rich limit [\autoref{fig:DFE_HFO}(b)], as follows directly from the $+\mu_{\mathrm{O}}$ contribution in \autoref{eq:DFE}.
This shift is substantially smaller than the approximately 5.6~eV permitted by the stability limits of bulk HfO$_2$ alone~\cite{008ShiyouChen2021}, reflecting the restricted oxygen chemical potential range imposed by LSMO.
Within the present equilibrium model, adoption of an LSMO electrode therefore limits the overall variation in oxygen-vacancy formation energies across the accessible oxygen conditions.
The existence of oxygen-deficient manganites such as SrMnO$_{2.5}$~\cite{Shin/Rondinelli2024,Wang/Shin2018,Han/Shin/Galli2026} and reports of oxygen transfer between HfO$_2$ and an oxide electrode~\cite{Lee/Chae2024} provide physical support for a possible oxygen-buffering role of LSMO.

\begin{table}[t]
    \centering
    \caption{Calculated charge transition levels of the 3fold- and 
    4fold-coordinated oxygen vacancies in o-HfO$_2$. 
    All transition levels are referenced to the valence-band maximum (VBM).}
    \label{tab:CTL_VO}
     \renewcommand{\arraystretch}{1.2}
\begin{ruledtabular}
\begin{tabular}{c c c}
    Transition level & 3fold (eV) & 4fold (eV) \\
    \hline
    $\varepsilon(+2/+1)$ & 3.500 & 3.379 \\
    $\varepsilon(+1/0)$  & 4.053 & 3.703 \\
    $\varepsilon(0/-1)$  & 5.297 & 5.239 \\
    $\varepsilon(-1/-2)$ & 5.299 & 5.484 \\
\end{tabular}
\end{ruledtabular}
\end{table}

\subsection{Properties of interstitial and substitutional defects}

Next, we discuss La-, Sr-, and Mn-related interstitial and substitutional defects in o-HfO$_2$. Under O-rich conditions, the interstitial defects have formation energies approximately 2--7~eV higher than the corresponding Hf-site substitutional defects at the VBM, as shown in \autoref{fig:DFE_HFO}(a).
La exhibits the largest interstitial--substitutional formation-energy difference, approximately 7~eV at the VBM, which increases further with the Fermi level.
Under O-poor conditions, however, the interstitial formation energies decrease by approximately 2~eV, while the substitutional formation energies increase by approximately 1~eV relative to the O-rich values [\autoref{fig:DFE_HFO}(b)]. These opposing shifts favor Sr$_i$ over Sr$_{\text{Hf}}$ and Mn$_i$ over Mn$_{\text{Hf}}$ for Fermi levels below approximately 0.5 and 1~eV above the VBM, respectively.

We now focus on the substitutional defects, which exhibit predominantly acceptor-like behavior, with negatively charged states thermodynamically stable over most of the band gap [\autoref{fig:DFE_HFO}(a)].
Neutral or positively charged states are favored only for Fermi levels within approximately 1~eV above the VBM.
For La$_{\text{Hf}}$ and Sr$_{\text{Hf}}$, the acceptor-like behavior reflects the valence mismatch between Hf$^{4+}$ and the substituting La$^{3+}$ or Sr$^{2+}$ cations~\cite{061_C.Radehaus_2011,049_Alfred.Kersch_2017}.
These substitutions introduce deficiencies of one and two electrons, respectively, which can be compensated by electron addition. Accordingly, La$_{\text{Hf}}$ and Sr$_{\text{Hf}}$ prefer the $q=-1$ and $q=-2$ charge states, respectively, over most of the band gap.

\begin{figure*}[t]
  \centering
  \includegraphics[width=0.95\textwidth,trim=0 0cm 0 0cm,clip]{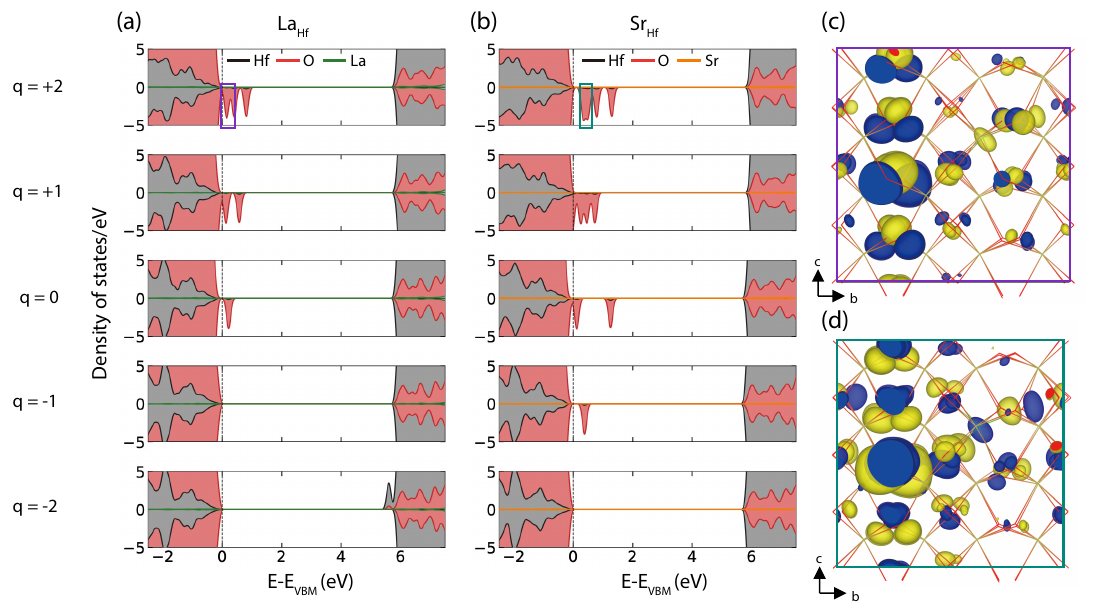}
  \caption{(a) and (b) Spin-resolved projected density of states (PDOS) of substitutional defects in o-HfO$2$ for charge states from $q=-2$ to $q=+2$: (a) La${_\text{Hf}}$ and (b) Sr${_\text{Hf}}$. (c) and (d) Wavefunctions corresponding to the O-derived defect state peaks highlighted in (a) and (b), respectively, for the $q=+2$ charge state: (c) La${_\text{Hf}}$ and (d) Sr${_\text{Hf}}$. In each wavefunction, yellow and blue represent the positive and negative phases, respectively.}
  \label{fig:DOS_substitution}
\end{figure*}

To examine this charge-compensation behavior, we analyzed the charge-state-dependent PDOS in \autoref{fig:DOS_substitution}(a).
In the neutral state, La$_{\text{Hf}}$ and Sr$_{\text{Hf}}$ exhibit one and two resolved O-derived PDOS peaks near the VBM, respectively.
Upon electron addition, the O-derived in-gap state of  La$_{\text{Hf}}$ disappears at $q=-1$, whereas  the in-gap states of Sr$_{\text{Hf}}$ disappear at $q=-2$.
This evolution is consistent with filling one and two oxygen holes associated with La$_{\text{Hf}}$ and Sr$_{\text{Hf}}$, respectively.
Conversely, electron removal introduces additional O-derived states near the VBM. For $q=+2$, the wavefunctions associated with the highlighted states extend over several O atoms surrounding the substitutional defects [\autoref{fig:DOS_substitution}(b,c)].

\begin{figure*}[t]
  \centering
  \includegraphics[width=0.85\linewidth,trim=0 0cm 0 0cm,clip]{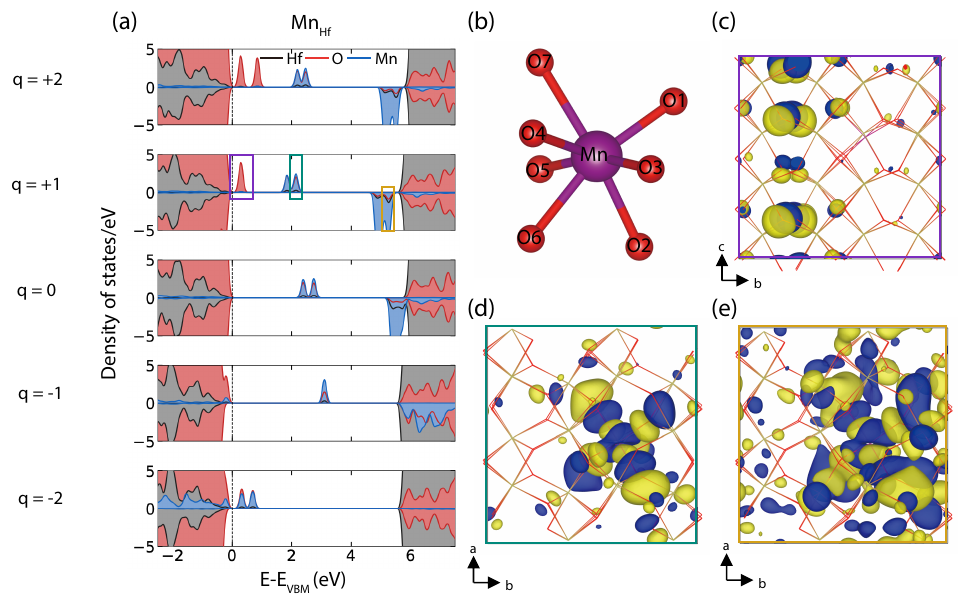}
  \caption{(a) Spin-resolved projected density of states (PDOS) of Mn$_{\text{Hf}}$ in o-HfO$_2$ for charge states from $q=-2$ to $q=+2$. (b) Atomic structure of Mn and the neighboring atoms at $q=-2$. (c)–(e) Wavefunctions corresponding to the defect-state peaks highlighted by the colored boxes in (a): (c) spin-up wavefunction corresponding to the peak highlighted by the purple box, (d) spin-up wavefunction corresponding to the peak highlighted by the teal box, and (e) spin-down wavefunction corresponding to the peak highlighted by the gold box. In each wavefunction, yellow and blue represent the positive and negative phases, respectively.}
  \label{fig:DOS:mnhf}
\end{figure*}

\begin{table*}[htbp]
\centering
\caption{Magnetic moments of Mn and the neighboring O atoms, the sum of the magnetic moments of the neighboring O atoms, and the magnetization of Mn${_\text{Hf}}$ in o-HfO$_2$ for different charge states. Magnetic moments and magnetization ($M$) values are in unit of $\mu_{\mathrm{B}}$.}
\label{tab:magnetic_moment}
\renewcommand{\arraystretch}{1.35}
\setlength{\tabcolsep}{8pt}
\begin{ruledtabular}
\begin{tabular}{r | cccccccc|cc}
$q$  & Mn  & O1  & O2  & O3  & O4  & O5  & O6 & O7 & Sum & $M$  \\
 \hline
 $+2$  & 3.488  & -0.638  & -0.113  & -0.148  & -0.071  & -0.086  & -0.126 & -0.013 & -1.195 & 1.26 \\
 $+1$  & 3.515  & -0.568  & -0.112  & -0.158  & -0.080  & -0.094  & -0.134 &  0.037 & -1.109 & 2.12 \\
 $0$   & 3.445  & -0.076  & -0.099  & -0.092  & -0.051  & -0.060  & -0.075 &  0.004 & -0.449 & 2.98 \\
 $-1$  & 3.960  & -0.043  & 0.045  & -0.046  & 0.021  & -0.033  & -0.034 & 0.01 & -0.080 & 3.90 \\
 $-2$  & 4.662  & 0.013  & 0.016  & 0.015  & 0.014  & 0.013  & 0.017 & 0.01 & 0.098 & 4.81 \\
 \end{tabular}
 \end{ruledtabular}
 \end{table*}

Unlike La$_{\text{Hf}}$ and Sr$_{\text{Hf}}$, Mn$_{\text{Hf}}$ forms in-gap states with predominantly Mn 3$d$ character in all surveyed charge states, with charge-dependent energies and occupations [\autoref{fig:DOS:mnhf}(a)].
Previous calculations of Mn-substituted HfO$_2$ likewise reported defect states with strong Mn 3$d$--O 2$p$ hybridization~\cite{087_Shuai.Zhao_2014}.
From the formation-energy diagrams in \autoref{fig:DFE_HFO}, we also find that Mn$_{\text{Hf}}$ has lower formation energies than La$_{\text{Hf}}$ and Sr$_{\text{Hf}}$ when the Fermi level exceeds approximately 3~eV above the VBM.
The thermodynamically stable charge state of Mn$_{\text{Hf}}$ becomes progressively more negative with increasing Fermi level, with the $(+2/0)$, $(0/-1)$, and $(-1/-2)$ transition levels located at 0.242, 1.076, and 2.22~eV above the VBM, respectively (\autoref{fig:DFE_HFO}).
Calculations for Mn-containing systems can converge to different magnetic states depending on the initial magnetic moments~\cite{054_MR.Wenman_2018}. We therefore tested multiple initial moment configurations on Mn and the seven neighboring O atoms (O1--O7) shown in \autoref{fig:DOS:mnhf}(b).
For each charge state, the lowest-energy magnetic configuration among those tested was selected for the subsequent analysis.

The wavefunctions of Mn$_{\text{Hf}}$ reveal a localized deep defect state and more spatially extended states near the band edges.
The O-derived state near the VBM extends over O atoms relatively far from Mn [\autoref{fig:DOS:mnhf}(c)], resembling the O-derived states of La$_{\text{Hf}}$ and Sr$_{\text{Hf}}$ [\autoref{fig:DOS_substitution}(b,c)].
The deep Mn-derived state in \autoref{fig:DOS:mnhf}(d) is strongly localized around Mn, whereas the state near the CBM in \autoref{fig:DOS:mnhf}(e) is more spatially extended.

At $q=-2$, which is stable over a wide Fermi-level interval (\autoref{fig:DFE_HFO}), the Mn magnetic moment is 4.66~$\mu_{\mathrm{B}}$ (\autoref{tab:magnetic_moment}), consistent with a high-spin $d^5$ configuration and Mn$^{2+}$-like character.
Removing electrons sequentially from $q=-2$ to $q=+2$ reduces the total magnetization ($M$) from 4.81 to 1.26~$\mu_{\mathrm{B}}$, or approximately 0.9~$\mu_{\mathrm{B}}$ per electron, consistent with preferential removal from the majority-spin channel.
The O-atom contribution becomes particularly pronounced from $q=0$ to $q=+1$: the Mn moment increases by only 0.07~$\mu_{\mathrm{B}}$, whereas the summed moment of O1--O7 becomes more negative by 0.66~$\mu_{\mathrm{B}}$, with O1 exhibiting the largest change.
From $q=+1$ to $q=+2$, $M$ decreases by 0.86~$\mu_{\mathrm{B}}$, whereas the combined local moment of Mn and O1--O7 decreases by only 0.11~$\mu_{\mathrm{B}}$, indicating that these local moments do not fully account for the spin response.

Electron removal also changes the local coordination environment of Mn.
As shown in \autoref{fig:DOS:mnhf}(b), Mn is sevenfold coordinated at $q=-2$. Upon electron removal, the Mn–O7 distance increases from 2.39~\AA{} at $q=-2$ to 2.77~\AA{} at $q=+2$, placing O7 outside the first coordination shell and leaving Mn sixfold coordinated.
The charge-state-dependent Mn--O bond lengths are provided in Section S4. 
These results suggest that, unlike the simpler charge compensation of La$_{\text{Hf}}$ and Sr$_{\text{Hf}}$, the charge-state evolution of Mn$_{\text{Hf}}$ involves electronic and magnetic responses of both Mn and neighboring O atoms, consistent with previous reports of O-atom magnetic contributions in Mn-doped HfO$_2$~\cite{086_DH.Hoat_2024}.

\begin{figure}[t]
\centering
\includegraphics[
width=\linewidth,
trim=0 0cm 0 0cm,
clip
]{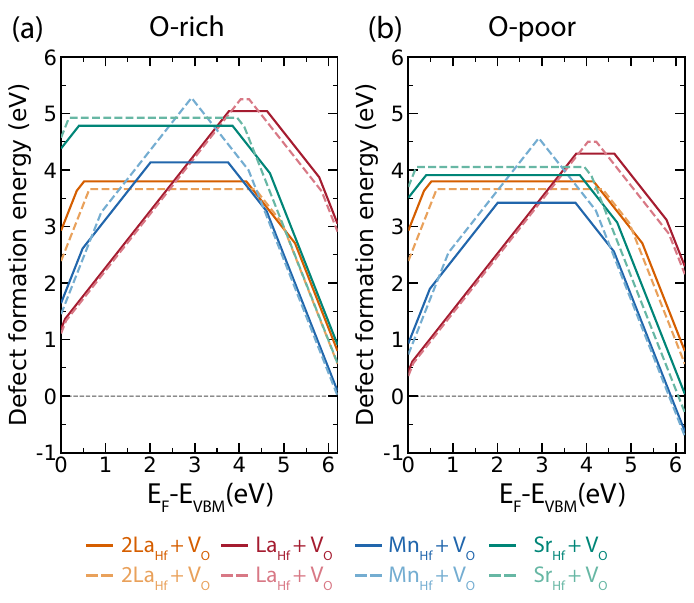}
\caption{Defect formation energy diagrams of paired defects involving oxygen vacancies and La, Mn, and Sr substitutional defects under (a) O-rich and (b) O-poor conditions. Solid and dashed lines represent 3fold and 4fold oxygen vacancy sites, respectively.}
\label{fig:paired}
\end{figure}

Additionally, we investigate defect complexes formed between substitutional defects (La${_\text{Hf}}$, Sr${_\text{Hf}}$, and Mn${_\text{Hf}}$) and oxygen vacancies to further examine the charge-compensation mechanisms discussed above. For La, we consider both La${_\text{Hf}}$–$V{_\text{O}}$ and 2La${_\text{Hf}}$–$V{_\text{O}}$ complexes to examine charge compensation involving one or two La substitutions, as shown in \autoref{fig:paired}. The reported HfO$_2$/LSMO band alignment places the Fermi level near mid-gap~\cite{084_Markus_Hellenbrand_2026}, corresponding to approximately 3.1 eV above the VBM for the calculated o-HfO$2$ band gap of 6.2 eV. At this Fermi level, $V_{\text{O}}^{2+}$ at the 3fold site, La$_{\text{Hf}}^{-}$, and Mn$_{\text{Hf}}^{2-}$ are the preferred charge states and configurations. These results suggest that one $V_{\text{O}}^{2+}$ can be charge-compensated by either one Mn$_{\text{Hf}}^{2-}$ or two La$_{\text{Hf}}^{-}$ defects. Previous calculations have also shown that La doping promotes oxygen-vacancy formation and that La$_{\text{Hf}}$ can form defect complexes with oxygen vacancies~\cite{070_YC.Zhou_2019}.

As shown in \autoref{fig:paired}, La- and Mn-containing defect complexes exhibit thermodynamic competition depending on the Fermi-level position and oxygen chemical potential. Under O-rich conditions, the most stable complex changes from La${_\text{Hf}}$–$V{_\text{O}}$ to 2La${_\text{Hf}}$–$V{_\text{O}}$ and then to Mn${_\text{Hf}}$–$V{_\text{O}}$ as the Fermi level increases from the VBM toward the CBM. Under O-poor conditions, the preferred complex instead changes directly from La${_\text{Hf}}$–$V{_\text{O}}$ to Mn${_\text{Hf}}$–$V{_\text{O}}$. Interestingly, unlike the other defect complexes, the formation energy of 2La${_\text{Hf}}$–$V{_\text{O}}$ exhibits a negligible change between O-rich and O-poor conditions due to the near cancellation of the chemical-potential contributions. To evaluate the interactions between substitutional defects and oxygen vacancies, we also calculated the binding energies using \autoref{eq:binding}. All defect complexes exhibit negative binding energies, with the 2La$_{\text{Hf}}$–$V{_\text{O}}$ complex showing the most negative value of $-1.683$ eV. The binding energies of La${_\text{Hf}}$–$V{_\text{O}}$, Mn${_\text{Hf}}$–$V{_\text{O}}$, and Sr${_\text{Hf}}$–$V{_\text{O}}$ are $-1.018$, $-1.200$, and $-1.260$ eV, respectively. These negative binding energies indicate that defect-complex formation is energetically favorable relative to the isolated defects. These results indicate that although both La- and Mn-containing complexes are energetically favorable relative to isolated defects, their relative formation energies depend strongly on the Fermi-level position and oxygen chemical potential.


\begin{figure*}[t]
  \centering
  \includegraphics[
    width=0.95\linewidth,
    trim=0 0cm 0 0cm,
    clip
  ]{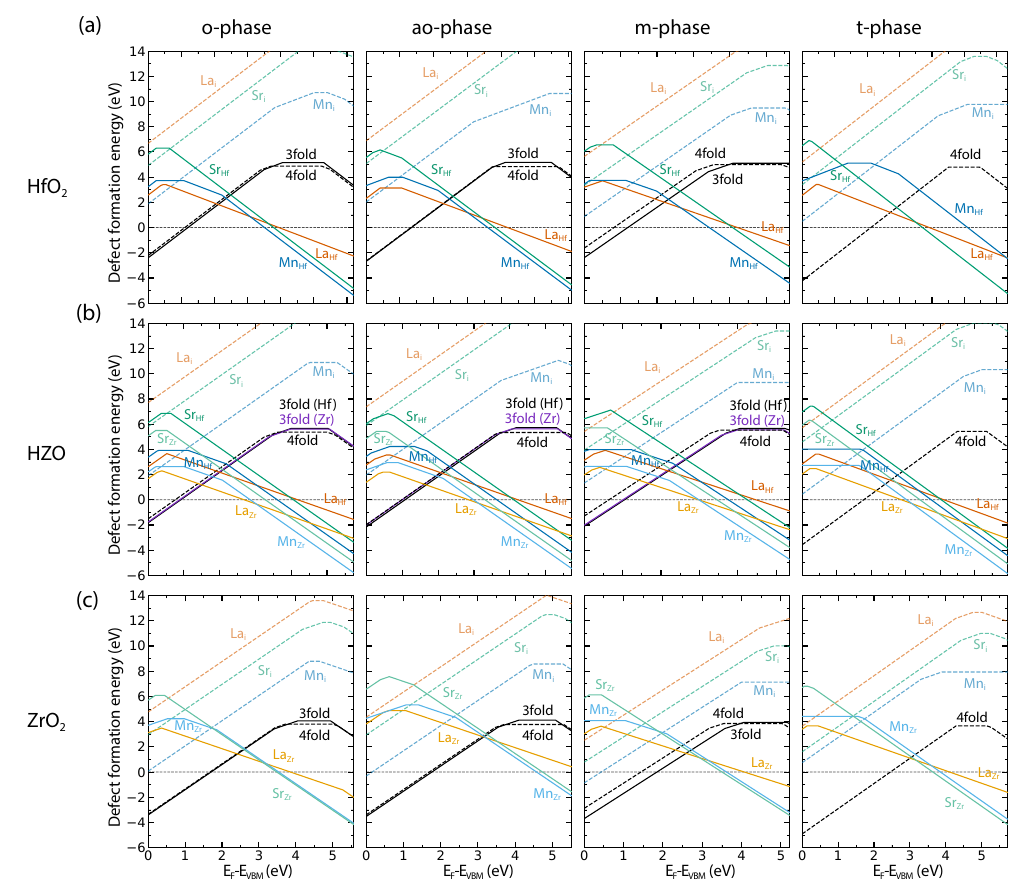}
  \caption{Defect formation energy diagrams of oxygen vacancies and La, Sr, and Mn interstitial and substitutional defects under O-poor conditions for the o-, ao-, m-, and t-phases of (a) HfO$_2$, (b) HZO, and (c) ZrO$_2$}
  \label{fig:All_formation}
\end{figure*}

\subsection{Defect Formation Energies across HfO$_2$, HZO, and ZrO$_2$ Phases}

We extend the analysis to other phases of HfO$_2$ and to HZO and ZrO$_2$ to examine the effects of host composition and crystal phase on defect formation. Under O-poor conditions, oxygen vacancies show similar charge-state and site-preference trends across the three hosts within a given crystal phase in \autoref{fig:All_formation}, consistent with the similar chemical properties of Hf and Zr~\cite{012_WK_Choi_2007}. In contrast, the absolute formation energies of LSMO-derived defects and the relative stability of substitutional defects vary substantially among the hosts. The corresponding defect formation-energy diagrams under O-rich conditions are provided in Section S2.

For all three hosts, $V_{\text{O}}^{0}$ generally favors the 4fold site over the 3fold site in the o-, ao-, and m-phases. This preference has been attributed to the energy increase associated with an occupied antibonding state formed upon oxygen removal at the 3fold site~\cite{089_Giuliana.Di.Martino_2026}. For positively charged vacancies, the formation-energy difference between the 3fold and 4fold sites is approximately 0.7--0.8~eV in the m-phase, being larger than in the o- and ao-phases. Across all hosts and phases, interstitial formation energies increase in the order Mn$_i$, Sr$_i$, and La$_i$. Interstitial formation energies in the t-phase are approximately 1--2~eV lower than those in the other phases.

Unlike the other HfO$_2$ phases, the m-phase has almost no Fermi-level interval in which negatively charged oxygen vacancies are thermodynamically stable [\autoref{fig:All_formation}(a)]. Across all four phases, interstitial defects can become more stable than the corresponding substitutional defects for Fermi levels below approximately 1.5~eV above the VBM. At Fermi levels more than approximately 3.5~eV above the VBM, Sr$_{\text{Hf}}$ is the most stable substitutional defect in the t-phase, whereas Mn$_{\text{Hf}}$ is favored in the o-, ao-, and m-phases.

In HZO, the local Hf/Zr environment affects the stability of both oxygen vacancies and substitutional defects. For neutral oxygen vacancies, the 3fold (Zr) configuration is more stable than the 3fold (Hf) configuration in the o-, ao-, and m-phases.
This preference has been attributed to a smaller O bonding contribution in the 3fold (Zr) configuration than in the 3fold (Hf) configuration~\cite{089_Giuliana.Di.Martino_2026}.
Substitutional defects show a more pronounced Zr-site preference: under the adopted chemical potentials, formation energies are approximately 1~eV lower at Zr sites than at Hf sites in all four phases [\autoref{fig:All_formation}(b)].
Zr-site substitution also consistently produces smaller displacements of the surrounding O atoms than Hf-site substitution across the examined dopants and phases. For example, for Mn substitution in the o-phase at $q=-2$, the average O displacement decreases from 0.102~\AA{} at the Hf site to 0.087~\AA{} at the Zr site, suggesting that reduced local distortion may contribute to the lower formation energy at the Zr site.

Under O-poor conditions, oxygen vacancies and La, Sr, and Mn interstitials have lower formation energies in ZrO$_2$ than in HfO$_2$ and HZO. The wider O chemical-potential range of ZrO$_2$/LSMO extends to $\Delta\mu_{\mathrm{O}}=-2.26$~eV [\autoref{fig:Chemical_potential}(d)], favoring oxygen-vacancy formation. This limit is obtained by considering metastable ZrO$_2$/LSMO conditions, where SrZrO$_3$ and La$_2$Zr$_2$O$_7$ are excluded from the competing phases based on kinetic considerations. The corresponding $V_{\text{O}}$ formation energies are approximately 0.7 and 1.4~eV lower than in HfO$_2$ and HZO, respectively.
For example, the formation energy of Mn$_i$ in m-ZrO$_2$ becomes negative for Fermi levels below approximately 0.5~eV above the VBM. In ZrO$_2$, interstitial defects can be more stable than the corresponding substitutional defects up to approximately 2~eV above the VBM, an interval approximately 0.5~eV wider than in HfO$_2$ and HZO.
For Fermi levels more than approximately 3~eV above the VBM, Sr$_{\text{Zr}}$ is the most stable substitutional defect in the o-, m-, and t-phases, whereas Mn$_{\text{Zr}}$ is favored in the ao-phase.

\section{Conclusion}

We demonstrate that the thermodynamic stability of LSMO narrows the accessible O chemical-potential range governing defect formation across the surveyed phases of HfO$_2$, HZO, and ZrO$_2$. For HfO$_2$, these constraints reduce the variation in oxygen-vacancy formation energy between the O-rich and O-poor limits from approximately 5.6 eV for the host oxide alone to approximately 1.5 eV. Electronic structure analysis further reveals that the defect states of $V_{\text{O}}$ are sensitive to the local coordination environment and structural relaxation. In o-HfO$_2$, La$_{\text{Hf}}$ and Sr$_{\text{Hf}}$ substitutions exhibit charge compensation associated with their respective one- and two-electron deficiencies, whereas Mn$_{\text{Hf}}$ accommodates charge-state changes through electronic and magnetic responses involving Mn and neighboring O atoms. Defect-complex calculations reveal negative binding energies ranging from $-1.02$ to $-1.68$ eV, indicating favorable interactions between substitutional defects and oxygen vacancies. The La- and Mn-containing complexes exhibit thermodynamic competition depending on the Fermi-level position and oxygen chemical potential. Near mid-gap, the 2La$_{\text{Hf}}$–$V_{\text{O}}$ complex is favored under O-rich conditions, whereas the Mn$_{\text{Hf}}$–$V{_\text{O}}$ complex is favored under O-poor conditions. Extending the analysis to HZO and ZrO$_2$ reveals that defect formation energies and relative stabilities depend strongly on host composition and crystal phase, despite broadly similar charge-state-dependent behavior of oxygen vacancies. Taken together, these findings establish how local defect chemistry and electrode-imposed chemical-potential constraints jointly govern defect formation and charge compensation in HfO$_2$-based oxides.

\FloatBarrier

\appendix


\FloatBarrier \bibliography{cite}

\end{document}



\noindent
{\bfseries Supplementary Information for}\\

\vspace{-0.3cm}

\noindent
{\bfseries\boldmath
Defect Chemistry of Hafnia-Based Ferroelectrics under Oxide-Electrode Constraints
}

\vspace{0.5cm}

\noindent
Sehyeok Jeon$^{1}$,
Aloysius Soon$^{1,\dagger}$,
and Yongjin Shin$^{2,*}$,

\vspace{0.3cm}

\renewcommand{\baselinestretch}{0.90}\normalsize

$^1$\textit{\footnotesize
Department of Materials Science and Engineering,
Yonsei University,
Seoul 03722,
South Korea}\\

$^2$\textit{\footnotesize
Department of Semiconductor Convergence Engineering,
Dankook University,
Yongin-si 16890,
Gyeonggi-do,
South Korea}\\

$^\dagger${\small Email: aloysius.soon@yonsei.ac.kr}\\
$^*${\small Email: yongjin.shin@dankook.ac.kr}\\

\vspace{0.75cm}


\renewcommand{\baselinestretch}{0.95}\normalsize

{
\hypersetup{linkcolor=black}
\tableofcontents
}

\renewcommand{\baselinestretch}{1.2}\normalsize

\newpage

\section{Crystallographic data for HfO$_2$}

\begin{table}[H]
\caption{\label{table:ohfo2}
Crystallographic data for orthorhombic HfO$_2$ in the $Pca2_1$ space group (No. 29), obtained by total-energy optimization with the PBEsol functional. Lattice parameters are $a$ = 5.2050~\AA, $b$ = 4.9978~\AA, $c$ = 5.0188~\AA, $\alpha = \beta = \gamma = 90^\circ$, and the unit-cell volume ($V$) is 130.5578~\AA$^3$.
}
\centering
\vspace{0.3em}

\setlength{\tabcolsep}{12pt}
\begin{tabular}{l|c|ccc}
\hline\hline
Atom & Site & $x$ & $y$ & $z$ \\
\hline\Tstrut
Hf1  & 4a  & 0.53389 & 0.26827 & 0.24260 \\
O1   & 4a  & 0.86878 & 0.07319 & 0.37878 \\
O2   & 4a  & 0.23670 & 0.45860 & 0.49162 \\
\hline\hline
\end{tabular}
\end{table}

\begin{table}[H]
\caption{\label{table:aohfo2}
Crystallographic data for antipolar orthorhombic HfO$_2$ in the $Pbca$ space group (No. 61), obtained by total-energy optimization with the PBEsol functional. Lattice parameters are $a$ = 5.0242~\AA, $b$ = 5.1949~\AA, $c$ = 9.9436~\AA, $\alpha = \beta = \gamma = 90^\circ$, and the unit-cell volume ($V$) is 259.5306~\AA$^3$.
}
\centering
\vspace{0.3em}

\setlength{\tabcolsep}{12pt}
\begin{tabular}{l|c|ccc}
\hline\hline
Atom & Site & $x$ & $y$ & $z$ \\
\hline\Tstrut
Hf1  & 8c  & 0.24559 & 0.11542 & 0.53510 \\
O1   & 8c  & 0.00244 & 0.02286 & 0.23867 \\
O2   & 8c  & 0.12580 & 0.71054 & 0.87351 \\
\hline\hline
\end{tabular}
\end{table}

\begin{table}[H]
\caption{\label{table:mhfo2}
Crystallographic data for monoclinic HfO$_2$ in the $P2_1/c$ space group (No. 14), obtained by total-energy optimization with the PBEsol functional. Lattice parameters are $a$ = 5.0735~\AA, $b$ = 5.1603~\AA, $c$ = 5.1640~\AA, $\alpha = \gamma = 90^\circ$, $\beta = 99.5813^\circ$, and the unit-cell volume ($V$) is 135.2159~\AA$^3$.
}
\centering
\vspace{0.3em}

\setlength{\tabcolsep}{12pt}
\begin{tabular}{l|c|ccc}
\hline\hline
Atom & Site & $x$ & $y$ & $z$ \\
\hline\Tstrut
Hf1  & 4e  & 0.72236 & 0.04298 & 0.79280 \\
O1   & 4e  & 0.55328 & 0.75853 & 0.51794 \\
O2   & 4e  & 0.07250 & 0.15980 & 0.16183 \\
\hline\hline
\end{tabular}
\end{table}

\begin{table}[H]
\caption{\label{table:thfo2}
Crystallographic data for tetragonal HfO$_2$ in the $P4_2/nmc$ space group (No. 137), obtained by total-energy optimization with the PBEsol functional. Lattice parameters are $a$ = 3.5513~\AA, $b$ = 3.5513~\AA, $c$ = 5.1295~\AA, $\alpha = \beta = \gamma = 90^\circ$, and the unit-cell volume ($V$) is 64.6935~\AA$^3$.
}
\centering
\vspace{0.3em}

\setlength{\tabcolsep}{12pt}
\begin{tabular}{l|c|ccc}
\hline\hline
Atom & Site & $x$ & $y$ & $z$ \\
\hline\Tstrut
Hf1  & 2a  & 0.75000 & 0.25000 & 0.75000 \\
O1   & 4d  & 0.25000 & 0.25000 & 0.54324 \\
\hline\hline
\end{tabular}
\end{table}

\begin{table}[H]
\caption{\label{table:chfo2}
Crystallographic data for cubic HfO$_2$ in the $Fm\bar{3}m$ space group (No. 225), obtained by total-energy optimization with the PBEsol functional. Lattice parameters are $a$ = 5.0191~\AA, $b$ = 5.0191~\AA, $c$ = 5.0191~\AA, $\alpha = \beta = \gamma = 90^\circ$, and the unit-cell volume ($V$) is 126.4360~\AA$^3$.
}
\centering
\vspace{0.3em}

\setlength{\tabcolsep}{12pt}
\begin{tabular}{l|c|ccc}
\hline\hline
Atom & Site & $x$ & $y$ & $z$ \\
\hline\Tstrut
Hf1  & 4a  & 0.00000 & 0.00000 & 0.00000 \\
O1   & 8c  & 0.25000 & 0.25000 & 0.25000 \\
\hline\hline
\end{tabular}
\end{table}

\clearpage

\section{Defect formation energy diagrams for HfO$_2$, HZO, and ZrO$_2$ under O-rich conditions}

\begin{figure*}[htbp]
\centering

\includegraphics[
    width=0.99\textwidth
]{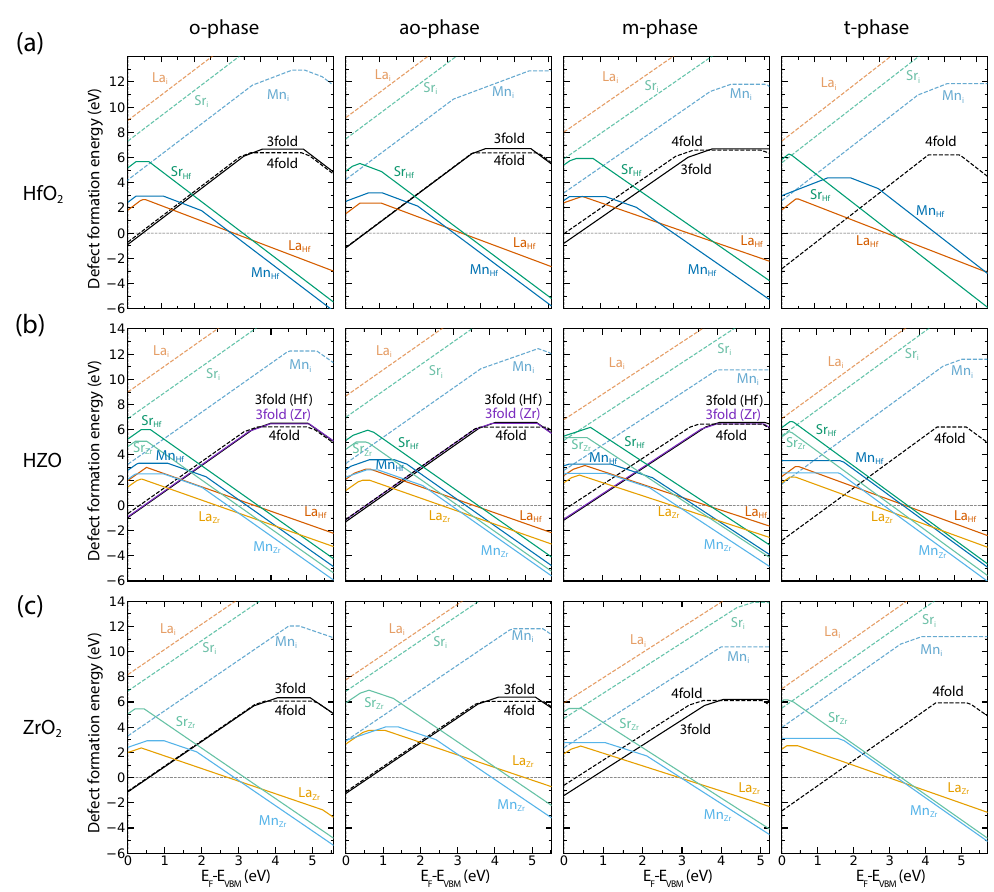}

\caption{\label{fig:S_O_rich_DFE}
Defect formation energy diagrams of oxygen vacancies and La, Sr, and Mn interstitial and substitutional defects under O-rich conditions for the orthorhombic (o), antipolar orthorhombic (ao), monoclinic (m), and tetragonal (t) phases of (a) HfO$_2$, (b) HZO, and (c) ZrO$_2$.
}

\end{figure*}

\clearpage


\section{Chemical potential constraints and competing phases}
\label{sec:chemical_potential}

The competing phases of LMO, SMO, and LSMO are summarized in \autoref{table:competing_phases}. As mentioned in the main text, the stability window of LSMO, taking into account the constraints required to prevent the formation of competing phases, is given as follows. 
\[
\begin{aligned}
\frac{2}{3}\Delta\mu_{\mathrm{La}}
+\frac{1}{3}\Delta\mu_{\mathrm{Sr}}
+\Delta\mu_{\mathrm{Mn}}
+3\Delta\mu_{\mathrm{O}}
&= \Delta H_f(\mathrm{LSMO}),
\\
\Delta\mu_{\mathrm{La}},\;
\Delta\mu_{\mathrm{Sr}},\;
\Delta\mu_{\mathrm{Mn}},\;
\Delta\mu_{\mathrm{O}}
&\leq 0,
\\
\Delta\mu_{\mathrm{Mn}} + \Delta\mu_{\mathrm{O}}
&\leq \Delta H_f(\mathrm{MnO}), \\
3\Delta\mu_{\mathrm{Mn}} + 4\Delta\mu_{\mathrm{O}}
&\leq \Delta H_f(\mathrm{Mn_3O_4}), \\
\Delta\mu_{\mathrm{Mn}} + 2\Delta\mu_{\mathrm{O}}
&\leq \Delta H_f(\mathrm{MnO_2}), \\
2\Delta\mu_{\mathrm{La}} + 3\Delta\mu_{\mathrm{O}}
&\leq \Delta H_f(\mathrm{La_2O_3}), \\
7\Delta\mu_{\mathrm{Sr}} + 4\Delta\mu_{\mathrm{Mn}}
+ 15\Delta\mu_{\mathrm{O}}
&\leq \Delta H_f(\mathrm{Sr_7Mn_4O_{15}}), \\
16\Delta\mu_{\mathrm{Sr}} + 8\Delta\mu_{\mathrm{Mn}}
+ 29\Delta\mu_{\mathrm{O}}
&\leq \Delta H_f(\mathrm{Sr_{16}Mn_8O_{29}}), \\
\Delta\mu_{\mathrm{Sr}} + 2\Delta\mu_{\mathrm{O}}
&\leq \Delta H_f(\mathrm{SrO_2}), \\
\Delta\mu_{\mathrm{Sr}} + \Delta\mu_{\mathrm{O}}
&\leq \Delta H_f(\mathrm{SrO}).
\end{aligned}
\]

Based on these conditions, the accessible chemical potential ranges for the coexistence of HfO$_2$/LSMO, HZO/LSMO, and ZrO$_2$/LSMO can be determined. However, for the ZrO$_2$/LSMO system, no thermodynamically stable region was found due to the formation of competing phases. Since the formation of competing solid phases can be kinetically hindered under realistic synthesis conditions, a constrained thermodynamic equilibrium was considered in this study~\cite{042_J.Neugebauer_2009,022_ChrisG.VandeWalle_2014,043_M.Scheffler_2003}. Specifically, ZrO$_2$/LSMO heterostructures can be formed at approximately $800^{\circ}\mathrm{C}$, whereas SrZrO$_3$ ($1300^{\circ}\mathrm{C}$) and La$_2$Zr$_2$O$_7$ ($1500^{\circ}\mathrm{C}$) generally require substantially higher synthesis temperatures~\cite{039_HongjieLuo_2009,040_H.Dai_2013,041_PhilippeLecoeur_2024}. Therefore, the constraints associated with SrZrO$_3$ and La$_2$Zr$_2$O$_7$ were excluded, resulting in a metastable chemical potential region for the ZrO$_2$/LSMO system.

\begin{table*}[htbp]

\caption{\label{table:competing_phases}
Competing phases considered in determining the chemical-potential
stability regions of the LMO, SMO, and LSMO.
}

\centering
\setlength{\tabcolsep}{10pt}

\begin{tabular}{ll}
\hline\hline

System & Competing phases \\

\hline\Tstrut

La--Mn--O &
La$_2$O$_3$,
MnO,
MnO$_2$,
Mn$_3$O$_4$
\\[0.5em]

Sr--Mn--O &
SrO,
SrO$_2$,
Sr$_7$Mn$_4$O$_{15}$,
Sr$_{16}$Mn$_8$O$_{29}$,
MnO,
MnO$_2$,
Mn$_3$O$_4$
\\[0.5em]

La--Sr--Mn--O &
MnO,
Mn$_3$O$_4$,
MnO$_2$,
La$_2$O$_3$,
Sr$_7$Mn$_4$O$_{15}$,
Sr$_{16}$Mn$_8$O$_{29}$,
SrO$_2$,
SrO
\\

\hline\hline
\end{tabular}

\end{table*}

\clearpage

\section{Local bonding environment of Mn$_{\text{Hf}}$ in o-HfO$_2$}
\label{sec:MnHf_structure}

\begin{table*}[htbp]

\caption{\label{table:MnO_bond}
Calculated Mn-O bond lengths (in \AA) around Mn$_{\text{Hf}}$ in o-HfO$_2$ for charge states from $q=-2$ to $q=+2$.
}

\centering

\setlength{\tabcolsep}{8pt}

\begin{tabular}{c c c c c c c c}

\hline\hline

Charge state
& O1
& O2
& O3
& O4
& O5
& O6
& O7 \\

\hline\Tstrut

$+2$
& 1.930
& 1.889
& 1.923
& 2.055
& 1.999
& 1.966
& 2.772 \\

$+1$
& 1.931
& 1.888
& 1.919
& 2.074
& 2.012
& 1.978
& 2.776 \\

$0$
& 1.932
& 1.887
& 1.913
& 2.097
& 2.026
& 1.991
& 2.776 \\

$-1$
& 1.934
& 2.146
& 1.972
& 2.423
& 2.039
& 2.063
& 2.582 \\

$-2$
& 2.066
& 2.229
& 2.183
& 2.324
& 2.156
& 2.294
& 2.389 \\

\hline\hline

\end{tabular}

\end{table*}


\section{Wavefunction of the 3fold oxygen site at $q=-1$}
\label{sec:vo}

\begin{figure*}[htbp]

\centering

\includegraphics[
    width=0.50\textwidth,
    trim=0 0 0 20,
    clip
]{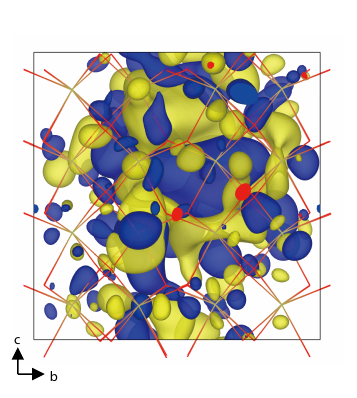}

\caption{\label{fig:S_vo}
Wavefunction associated with the peak immediately above the conduction band minimum (CBM) for the 3fold oxygen site in o-HfO$_2$ at the $q=-1$ charge state. Yellow and blue represent the positive and negative phases, respectively.
}

\end{figure*}

\clearpage

\bibliographystyle{apsrev4-2}
\bibliography{cite}